\documentclass[aps,prl,preprint,superscriptaddress]{revtex4}

\usepackage{graphics}
\usepackage{psfrag}
\usepackage{epsfig}
\usepackage{subfigure}
\usepackage{multirow}
\usepackage{epic}

\begin{document}

\preprint{LMU-ASC 13/05}

\title{Melting of Colloidal Molecular Crystals on Triangular Lattices}

\author{A. \v Sarlah$^{1}$\footnote{On leave from University of
  Ljubljana.}, T. Franosch$^{1,2}$, and E. Frey$^{2}$}

\affiliation{$^1$Hahn-Meitner-Institut, Abteilung Theorie, Glienicker
  Strasse 100, D-14109 Berlin, Germany \\
  $^2$Arnold Sommerfeld Center and CeNS, Department of Physics,
  Ludwig-Maximilians-Universit\"at M\"unchen, Theresienstrasse 37,
  D-80333 M\"unchen, Germany}

\date{\today}

\begin{abstract}
  The phase behavior of a two-dimensional colloidal system subject to
  a commensurate triangular potential is investigated. We consider the
  integer number of colloids in each potential minimum as rigid
  composite objects with effective discrete degrees of freedom. It is
  shown that there is a rich variety of phases including ``herring
  bone'' and ``Japanese 6 in 1'' phases. The ensuing phase diagram and
  phase transitions are analyzed analytically within variational
  mean-field theory and supplemented by Monte Carlo simulations.
  Consequences for experiments are discussed.
\end{abstract}


\maketitle

In the last decade it has been realized that soft materials can serve
as versatile model systems to study phenomena of condensed matter
physics.  In particular, two-dimensional (2D) systems of colloidal
particles interacting with light sources are ideal to mimic the
adsorption of atoms and molecules on atomic surfaces, vortices in
superconductors with periodic pinning arrays, and many other related
phenomena~\cite{berlinsky1978,pokrovsky1984,matsuda1996}.
Experimental studies on flat substrates~\cite{murray1987-zahn1999}
have beautifully confirmed the existence of a two-stage melting
process mediated by the successive unbinding of
dislocations~\cite{kosterlitz1973} and
disclinations~\cite{nelson1979}. If confined to a 1D periodic
potential 2D colloidal systems show even richer behavior with
re-entrant melting~\cite{wei1998} and novel phases such as the locked
floating solid~\cite{frey1999,baumgartl2004}. Results from analytical
theories have been complemented by a series of numerical simulations
(see e.g.
Refs.~\cite{loudiyi1992-1,chakrabarti1995-das2001,strepp2002}).

Recent experimental~\cite{brunner2002,mangold2003} and
theoretical~\cite{reichhardt2002,agra2004} investigations have studied
the effect of {\em 2D periodic potentials} on the phase behavior. In
particular, experiments on triangular light
lattices~\cite{brunner2002} with a stochiometry of three colloids per
site have motivated our research.  One observes that even at rather
low potential strength the colloids tend to group as {\em trimers}
forming an almost equilateral triangle, but with a quite significant
number of defects, i.e., groups of two or four particles. Upon
increasing the potential strength the number of defects decreases
rapidly such that they become unobservable already for moderate
intensities.  Then, the important low energy excitations are the
orientation of the trimers, regarded as rigid composite objects, with
respect to the lattice direction.  Due to the interaction of the
trimers long-range orientational order is expected for sufficiently
strong coupling.  Alignment of the trimers is observed
experimentally~\cite{brunner2002} as soon as the defect-free regime
takes over.  Interestingly, the same authors also observed a loss of
orientational long-range order at even higher intensities, which was
interpreted as re-entrant melting and seems to be confirmed by
computer simulations~\cite{reichhardt2002}.

In this Letter we derive phase diagrams for commensurate colloidal
systems in high external fields by analytic methods supplemented by
Monte-Carlo (MC) simulations. The key idea is to reduce the problem to
the low energy degrees of freedom by considering the integer number of
colloidal particles which gather in a single potential minimum as a
{\em rigid composite object}. Its shape is determined by the symmetry
of the lattice and the number of the constituting colloidal particles,
and its size by the interplay of interparticle repulsion and external
potential. Short-time orientational fluctuations close to the
potential minima are considered to be already averaged out leaving
only a discrete set of gross orientations. We will mainly focus on
{\em dimers} on a triangular lattice since they exhibit a rich phase
diagram and exhibit a series of intriguing phases. At the end of the
letter we will shortly report how the phase diagrams for trimers can
be obtained in a rather straightforward manner.

\begin{figure}[htb]
\begin{center}
  \epsfig{file=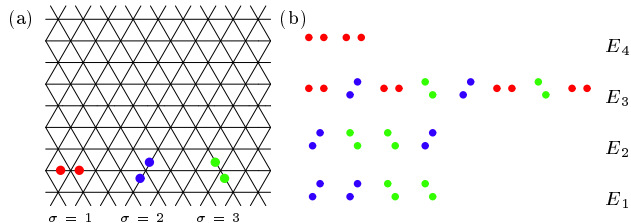}
\end{center}
\caption{The model system. (a) The triangular lattice of the 
  external field and the three orientational states of the dimers. (b)
  Four energy levels for the pair-wise interactions of dimers.}
\label{fig1}
\end{figure}

For an isolated dimer there are three equivalent orientational ground
states on a triangular lattice, denoted by $\sigma = 1,2,3$ (see
Fig.~\ref{fig1}).  
The effective interaction between the dimers results from the screened
Coulomb interaction between the constituent colloids, which is
short-ranged for the experimental conditions in
Ref.~\cite{brunner2002}. Therefore, it is appropriate to restrict the
effective dimer--dimer interaction to a nearest neighbors.
A pair of dimers can be in $3^2$ configurations corresponding to one
of four energy levels, with generic ordering $E_1 < E_2 < E_3 < E_4$;
see Fig.~\ref{fig1}. The relation between these model parameters and
the experimental control parameters like screening length and
potential strength can be worked out in detail~\cite{sarlah2004-2},
but is far too involved to be presented here. As a guidance there are
the following trends. Increasing the laser potential compresses the
dimers and thus by increasing the effective distances between
neighboring dimers reduces their interaction energies and hence the
values of $E_i$ and the spacings between them. This is the basic
mechanism underlying orientational melting. Lowering the screening
length effectively corresponds to increasing the potential strength.

The Hamiltonian of our model system reads
\begin{equation}
   {\cal H} = \sum_{\alpha = 1}^3 \sum_{\langle i,j \rangle_\alpha}
   \underline{\sigma}_i^\top \cdot  
   {\sf H}_{\alpha} \cdot \underline{\sigma}_j \, ,
\label{eqn1}
\end{equation}
where $\langle i,j \rangle_\alpha$ denotes the pair of nearest
neighbors whose bond-vector is parallel to the orientation of the
dimer state $\alpha$. $\underline{\sigma}_i$ is a vector
representation of the dimer configuration $\sigma_i$ at the site $i$:
$1 \mapsto \underline{\sigma} = (1,0,0)$, $2 \mapsto
\underline{\sigma} = (0,1,0)$, and $3 \mapsto \underline{\sigma} =
(0,0,1)$. For the energy matrices ${\sf H}_{\alpha}$ one finds in
terms of a permutation matrix
\begin{eqnarray}
        {\sf H}_1 &=& \left(\matrix{E_4 & E_3 & E_3 \cr
                                  E_3 & E_1 & E_2 \cr
                                  E_3 & E_2 & E_1}\right) , \quad
        {\sf P} = \left(\matrix{0 & 1 & 0 \cr
                                  0 & 0 & 1 \cr
                                  1 & 0 & 0}\right) \nonumber\\
        {\sf H}_2 &=& {\sf P}^\top {\sf H}_1 {\sf P}, 
        \quad {\sf H}_3 = {\sf P} {\sf H}_1 {\sf P}^\top \, .
\label{eqn2}
\end{eqnarray}

Since a shift of the global energy scale does not affect the phase
behavior only the three energy differences in units of $k_B T$
constitute dimensionless parameters. Thus one would be led to expect a
three-dimensional phase diagram. However, upon rewriting the
Hamiltonian in ``spin language'' it turns out that the parameter space
can actually be reduced to only two dimensions~\cite{sarlah2004-2}.
We find, omitting an additive constant
\begin{equation}
        {\cal H} = - K \sum_{\langle i,j \rangle} \delta_{\sigma_i,\sigma_j}
                - M \sum_{\alpha = 1}^3 \sum_{\langle i,j \rangle_\alpha}
                        \delta_{\sigma_i,\alpha}
        \delta_{\sigma_j,\alpha}  \, ,
\label{eqn3}
\end{equation}
where the new energy scales $K,M$ are given in terms of the direct
dimer interaction energies by $K = - E_1 + E_2$ and $M = E_1 - 2 E_2 +
2 E_3 - E_4$. This reduction greatly simplifies the analysis and
facilitates comparison with future experiments with the two generic
control parameters, screening length and laser intensity~\footnote{Both
  upon increasing the potential strength and the screening length the
  ratio $|M|/K$ becomes smaller.}.

For $M = 0$ the Hamiltonian reduces to a 3-state Potts model. The
critical properties for the 2D $q$-state Potts model for $q \geq 4$
can be determined exactly for ferromagnetic (FM) interactions, $K >
0$. The critical point is known rigorously for square, triangular, and
honeycomb lattices~\cite{baxter1989,wu1982}. There is sufficient
numerical evidence that the $q \geq 4$ solution also holds for $q = 3$
although an analytic proof is still missing. The FM--P (paramagnetic)
phase transition of the 3-state Potts model is continuous as suggested
by the assumed ``exact'' solution and corroborated by MC simulations
and renormalization group analysis~\cite{wu1982}. Interestingly, a
mean-field description gives qualitatively wrong results, since it
predicts a discontinuous transition. For negative exchange coupling $K
< 0$ the low temperature phase is an antiferromagnetic (AM) Potts
state. The results of renormalization group, series expansions, and MC
simulations for the transition on a triangular lattice appear to be
controversial as far as the nature of the transition is
concerned~\cite{wu1982}.

The phase diagram of dimers in the external potential forming a
triangular lattice as prescribed by a generalized Hamiltonian,
Eq.~(\ref{eqn3}), has a richer topology. Generically $K \ne 0$, and
one distinguishes between $K > 0$ and $K < 0$. In both cases there are
two dimensionless parameters, the ratio of the exchange couplings
$\tilde{M} = M/|K|$ and the normalized temperature $\tilde{T} = k_B
T/|K|$.

An exact analytic solution of the whole phase diagram, except for the
two Potts points, is difficult to obtain and one has to rely on
approximative and/or numerical methods. We start with a mean-field
(MF) analysis which allows us to determine the topology of the phase
diagram and thereby the symmetry of the order parameter in the various
low temperature phases. In particular, we employ a variational MF
approach where the full density matrix is approximated as a product of
single-site density matrices, $\rho(\{\sigma_i\}) = \prod_i
\rho_i(\sigma_i)$. By choosing site-independent density matrices the
order parameter of the FM phase has been found by minimizing the
variational free energy.  Appropriate generalizations have been
introduced to obtain AM phases and other ordered phases; details will
be presented elsewhere~\cite{sarlah2004-2}.
\begin{figure}
        \epsfig{file=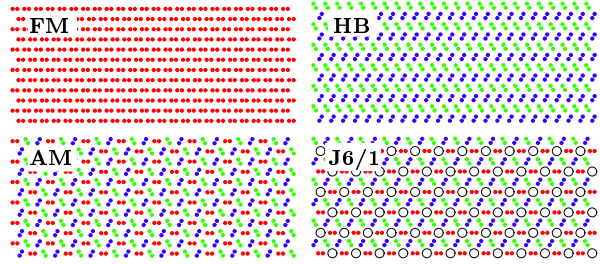}
\caption{Zero temperature ordered structures of 2D colloidal 
  dimers: (a) ferromagnetic (FM), (b) herring bone (HB), (c) Potts
  antiferromagnetic (AM), and (d) Japanese 6 in 1 (J6/1), structure.
  In the latter, the energy of the system does not depend on the
  orientation of dimers on sites denoted by circles, thus this
  structure is $3^{N/4}$ degenerate.}
\label{fig2}
\end{figure}

{F}rom the MF analysis we find the equation of state for the
respective order parameters, and in particular the phase boundaries.
It turns out that the variational free energy for colloidal dimers in
the P/FM or AM phase is identical to the corresponding result for the
Potts model, provided one substitutes the Potts exchange energy $K$ by
the effective coupling $ K \to K_{\mbox{\scriptsize eff}} = K + M /
3$. The critical temperatures for the P--FM and P--AM phase
transition are located at $k_B T_c = 3 K_{\mbox{\scriptsize eff}} / (2
\ln 2)$ and $k_B T_c = - 3 K_{\mbox{\scriptsize eff}} / (4 \ln 2)$,
respectively.

\begin{figure}[htb]
        \epsfig{file=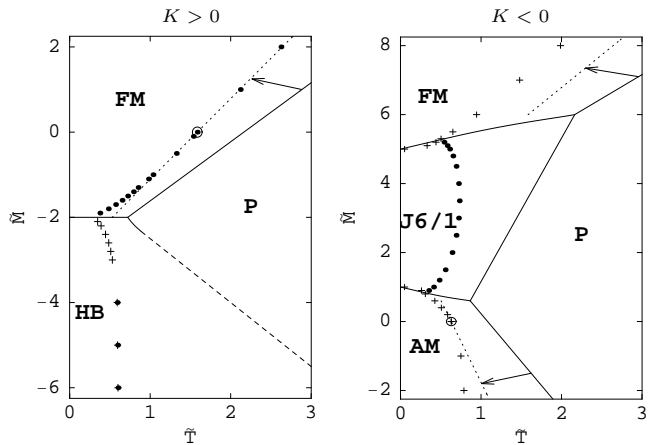}
\caption{Phase diagram for dimers. Solid/dashed lines
  represent discontinuous/continuous MF transitions. Dots and crosses
  denote the continuous and discontinuous transition points as
  obtained by MC; the two encircled symbols indicate the pure Potts
  transitions. The nature of the HB--P transition for $\tilde{M}<-4$
  is not clear which is indicated by a dot-cross symbol for the
  critical point. Dotted lines represent extrapolations of the Potts
  model critical point, see text.}
\label{fig3}
\end{figure}

The colloidal dimer Hamiltonian, Eq.~(\ref{eqn3}), allows for new
intriguing structures which are not realized within the Potts model.
We have investigated for {\em herring bone} (HB) and {\em Japanese 6
  in 1} (J6/1) structures -- a term borrowed from weaving patterns for
chainmailles --, see Fig.~\ref{fig2}. The corresponding order
parameters can be obtained by (numerically) minimizing the variational
free energies and the phase boundaries can be determined accordingly;
see Fig.~\ref{fig3}. Except for the HB--P transition for
$\tilde{M}<-7/3$ within MF theory all phase transitions are
discontinuous. The HB--P transition has already been studied in the
context of N$_2$ absorbates on graphite. A continuous anisotropic
planar rotor model analogous to our discrete version for $K=0$ yielded
similar results both within MF and MC
description~\cite{mouritsen1982,marx1994}.

\begin{figure}[htb]
        \epsfig{file=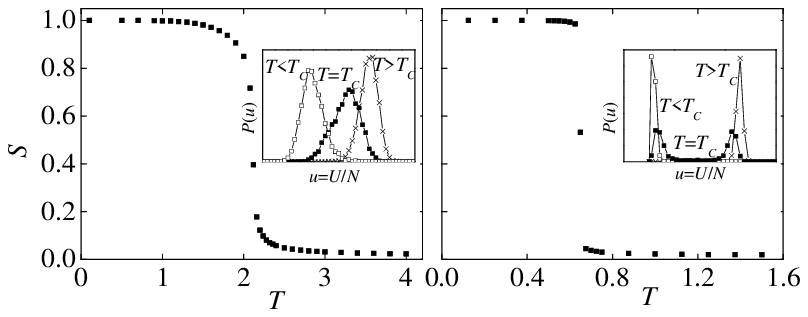,width=7.7cm}
\caption{Order parameter $S$ as a function of the reduced 
  temperature for the two FM--P transitions. Left: Potts-like
  continuous transition for $K,M > 0$. Right: the $\tilde{M}$-term
  driven first order transition for $K < 0, \tilde{M} > 5$. Inset:
  probability distributions of states with respect to the energy per
  dimer, $u = U/N$, below, above, and at the transition temperature.}
\label{fig4}
\end{figure}

In our case, the MF description is not expected to predict the correct
order of the transition as can be inferred by specializing to the
Potts model ($M = 0$). Nevertheless, the MF results should constitute
reasonable approximations for the actual phase behavior. In order
to gain further insight we have performed extensive MC simulations
of the
colloidal dimer Hamiltonian, Eq.~(\ref{eqn3}), using a standard
Metropolis procedure. The observables have been averaged over $\geq
5000$ statistically independent configurations after the system has
been equilibrated.  Every $1000$ MC cycles a new configuration
contributing to the averages is considered.  Simulations have been
performed on several system sizes, $N = L \times L$ and $L =
12,36,54,108$, and periodic boundary conditions have been employed to
mimic bulk-like conditions~\cite{sarlah2004-2}. The order parameter
characterizing the various phases has been chosen in analogy with the
Potts model, either defined on the whole system or suitable
sublattices,
\begin{equation}
  S = \frac{1}{N} 
      \biggl[ N_{\Sigma}  - \frac12 \sum_{\sigma \ne \Sigma}
      N_{\sigma} \biggr] \, .
\label{eqn5}
\end{equation}
Here, $N_\sigma$ is the number of dimers in state $\sigma$, in
particular, $N_{\Sigma}$ corresponds to majority orientation, and $N =
\sum_\sigma N_\sigma$ is the total number of the dimers on the whole
(sub)lattice. We have performed temperature sweeps for $K > 0$ and $K
< 0$ with a number of choices for $\tilde{M}$ starting from the
ordered structures. We have monitored the average order parameter and
energy, as well as their respective variance.  The transition
temperatures corresponding to melting of the ordered structures have
been determined by extrapolating the location of the maxima of the
heat capacity and susceptibility to the infinite system size. In order
to decide whether the respective phase transition is strongly
discontinuous or compatible with a continuous/weakly discontinuous
transition scenario we have evaluated Binder's cumulant and respective
probability distributions for the order parameter and
energy~\cite{sarlah2004-2}; both scenarios are exemplified in
Fig.~\ref{fig4}.

Figure~\ref{fig3} exhibits the phase diagram for the colloidal dimer
Hamiltonian; boundaries resulting from both MF description and MC
simulations are included in the Figure. The variational MF free energy
reproduces the topology of the phase diagram, although some of the
transitions become continuous by fluctuation effects. Quantitatively,
MF overestimates the transition temperatures with respect to
simulation results by a factor of order unity, a feature already
familiar from the Ising model~\cite{baxter1989}. Interestingly, the
discontinuous HB--P transition appears to saturate for large negative
values of $\tilde{M}$ in strong contrast to the corresponding MF
result~\footnote{Single-dimer flips of the HB ground state correspond
  to excitation energies $2K$ and $2|M|$ depending on the two possible
  new orientations of the dimer. For $-M \gg K$, $M$-type excitations
  are strongly suppressed.  However, the MF approach, in which
  generically the role of $M$ is to renormalize the energy $K$ in a
  linear way, misses that.}. For colloidal systems the regime
corresponding to $K > 0$ and $\tilde{M} < 0$ should be experimentally
accessible. A stringent test of our theory would be to experimentally
verify the predicted HB--P phase transition.

Fluctuation effects reduce the critical temperature of the Potts model
by a known factor~\cite{wu1982}. The variational MF free energy of the
colloidal dimer Hamiltonian for P/FM and AM structures is equivalent
to the MF Potts model, provided that the Potts exchange energy is
replaced by the effective coupling $K_{\mbox{\scriptsize eff}}$. This
analogy suggests an empirical improvement of the MF result for these
transitions by rescaling the transition temperatures with the same
respective correction factors. The result of this procedure is shown
as dashed lines in Fig.~\ref{fig3}. For $K > 0$ there is excellent
agreement with simulation results for temperatures $\tilde{T} > 1$.
For $K < 0$ the mapping gives an overall quantitative improvement of
the phase diagram. In particular, for $M = 0$ the well known Potts
solution is recovered.

We close with reporting our main results for {\em trimers} on a
triangular lattice. Due to the symmetry of the lattice, trimers can
only be in one of two orientational states, represented by Ising spins
$\sigma = \pm$. Hence the four configurational states for trimer pairs
contribute one out of the three possible interaction energies,
\begin{equation}
\epsfig{file=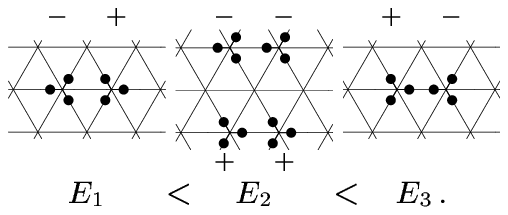}
\end{equation}
In contrast to the Ising model the interaction energies for the $(+-)$
and $(-+)$ configurations are different. The relation between the
model parameters and the control parameters of the experiments are
similar as for dimers. In particular, an increase of the potential
strength again leads to a reduction of the interaction energies
between trimers and thus favors orientational melting.

The central questions now are how many relevant energies determine the
phase behavior and what is the effective statistical model. Similar
considerations as for dimers would suggest that there are two relevant
energy scales resulting in a two-dimensional phase diagram.  Actually,
by a simple rearrangement the Hamiltonian can be cast in the form of a
Ising model with a single excitation energy $\Delta E = E_1 + E_3 - 2
E_2$~\cite{sarlah2004-2} corresponding to single trimer flips,
irrespective of the configurations of the neighboring trimers.
Depending on the sign of $\Delta E$ the ground state is either
``ferromagnetic'' with parallel $(++)$ (or equivalently $(--)$)
configurations or ``antiferromagnetic'' with an alternating
arrangement of $(-+)$ and $(+-)$ configurations. It is interesting to
note that it is the relative orientation of trimer pairs and not the
trimers itself which determines the ground state.

For the remainder of the analysis known {\em exact} results of the 2D
Ising model~\cite{baxter1989} can now be employed to obtain the phase
diagram.  Considering the parameters of the experimental set-up in
Ref.~\cite{brunner2002} we can identify the critical laser intensity,
$V_0=78~k_B T$~\cite{sarlah2004-2}, above which the orientational
order of trimers is lost. This result is consistent with experimental
observations, where the colloidal trimer system was found to be
orientationally ordered for $V_0=60~k_B T$ but disordered for
$V_0=110~k_B T$.  It would be interesting to extend the experimental
analysis to find the actual transition point and compare it with our
prediction.

\begin{acknowledgments}
  We thank C. Bechinger for discussions. A. \v S. acknowledges the
  support by the Humboldt Foundation.
\end{acknowledgments}

\end{document}